\documentclass[preprint,nofootinbib]{revtex4}%
\usepackage{amssymb}
\usepackage{amsfonts}
\usepackage{amsmath}
\usepackage{graphicx}%
\usepackage{xcolor}
\begin{document}
	\begin{titlepage}
\vskip1cm
\begin{center}
	{~\\[140pt]{ \LARGE {\textsc{Thermalization of horizon through asymptotic symmetry in three-dimensional massive gravity}}}\\[-20pt]}
\vskip2cm
\end{center}
\begin{center}
{M. R. Setare$^a$ \footnote{E-mail: rezakord@ipm.ir}\hspace{1mm} ,
A. Jalali$^a$ \footnote{E-mail: alijalali.alijalali@gmail.com}, Bibhas Ranjan Majhi$^b$ \footnote{E-mail: bibhas.majhi@iitg.ac.in}\hspace{1.5mm} \\
{\small {\em  {$^a$Department of Science,
Campus of Bijar, University of Kurdistan, Bijar, Iran.\\
$^b$Department of Physics, Indian Institute of Technology Guwahati, Guwahati 781039, Assam, India.
}}}}\\
\end{center}
\vspace{1cm}
Recently, black hole symmetries have been studied widely
 and it has been speculated that this procedure will lead to the deeper understanding of the black hole physics. Spontaneous symmetry breaking of the horizon symmetries is one of the very recent attempt to  clarify black hole thermal physics. In this work, we are going to investigate the same in three dimensional massive gravity, including higher order of Riemann tensor. We observe that the idea also works well in this gravitational theory, thereby providing stronger demand of the viability of this idea.
\vspace{0.8cm}
\end{titlepage}

\section{Introduction}	
Black holes are thermal objects \cite{Bekenstein:1973ur,Bardeen:1973gs,Hawking:1974sw} and the thermalization of horizon is meaningful only at the quantum level. But so far the microscopic description of the thermodynamics is still illusive. There has been a surge of works with the common believe that the asymptotic symmetries of the spacetime can shed some light. The symmetries has been investigated on both surfaces of a black hole spacetime: at the asymptotic infinity \cite{Brown:1986nw} as well as near the horizon \cite{Carlip:1998wz,Carlip:1999cy,Majhi:2011ws,Majhi:2017fua,Majhi:2012tf,Bhattacharya:2018epn} (also see \cite{Majhi:2012nq} for an extensive list). Among different asymptotic symmetries, recently the works of Bondi-Metzner-Sachs (BMS) \cite{b} attracted a lot of attention. BMS is originally applicable at the asymptotic infinity. The same idea has also been extended towards the horizon as well \cite{Donnay:2020yxw,Donnay:2018ckb,Maitra:2018saa,Maitra:2019eix}. Recently similar approach has been put forward for timelike surface also \cite{Maitra:2020ycr}.

Usually the BMS symmetry is accompanied by parameters like supertranslation alone or both supertranslation and superrotation, depending upon the nature of spacetime. These parameters in general modifies the black hole solution and thereby changes the macroscopic parameters of the spacetime. Although boundary conditions on the metric coefficients do respect the asymptotic structure of the metric, but they modifies the quantities of black hole like mass, charge, angular momentum, etc. This can be interpreted as the spontaneous symmetry braking of the metric solution and thereby introduces the concepts of emerged Goldstone modes \cite{Eling:2016xlx,Eling:2016qvx,Maitra:2019eix}. Here we are not going to repeat the  discussion about this idea elaborately. One can see \cite{Maitra:2019eix} for more explanatory introduction of Goldstone mode concepts in the context of black holes. A general believe is that these modes may help to revel the underlying microscopic origin of horizon thermodynamics. Recently the dynamics of these modes has been investigated in \cite{Maitra:2019eix} within the Einstein's gravity in the near horizon surface. It is observed that the time components of the Fourier modes behave like one dimensional inverse harmonic oscillator (IHO) which in turn suffers a local instability and thereby providing a temperature upon quantization. It is therefore igniting a hope towards the aforesaid general believe.

One now needs to search for the generality of the observation done in \cite{Maitra:2019eix}. Question is whether Goldstone modes can provide explanation for temperature to the horizon if one extend the analysis beyond Einstein gravity.  Having this motivation, in this paper we will concentrate on massive theory of gravity in $(2+1)$ dimensions by Bergshoeff, Hohm and Townsend (BHT). The general relativity in three spacetime dimensions has become an increasingly popular model in which people explore the foundations of classical and quantum gravity \cite{Steif:1995zm,Chan:1994rs,Wald:1984rg}.  Although (2+1)-dimensional gravity has been
widely recognized as a useful laboratory for studying conceptual issues, it has been widely believed that the model is too physically unrealistic to give much insight into real gravitating systems in 3+1 dimensions. However, in 1992 it has been shown that (2+1)-dimensional gravity has a black hole solution that is known as the BTZ black hole \cite{Banados:1992wn}. Here we are going to study this theory following the steps which are identical to those adopted in \cite{Maitra:2019eix}.

We first impose the required conditions on the metric solution of BHT gravity theory which preserves the near horizon structure to find the diffeomorphism parameters. Using this the modified metric is being then obtained. We then proceed to find the solution of this parameter, identified as Goldstone field, which makes the modified metric to be a solution of BHT gravity action on the horizon. This leads to an equation for the time part of the modes of the parameter which is that of a one dimensional IHO. A semiclassical quantization scheme of IHO, well established
in literature, leads to a concept of temperature. We find that the temperature of the zeroth mode is given by the Hawking expression for this black hole. This once again bats for the arguments that the asymptotic symmetries of horizon may carry the information about the microscopic degrees of freedom for black hole thermalization.

The organization of the paper as follows. In the next section we briefly introduce the BHT theory and its black hole solution. Section \ref{Bibhas3} is devoted for finding the diffeomorphism symmetries and the dynamics of the Goldstone parameter. Semi-classical analysis of this is being done in section \ref{Bibhas4} which shows that the thermal nature of horizon is originated from the modes of the Goldstone field. Next section discusses the BTZ limit of this analysis and finally we conclude in section \ref{Bibhas6}.

\section{BHT massive gravity and black hole}
	Among massive gravity models in three dimensions, a well known one is that of the new massive gravity (NMG) or BHT model \cite{Bergshoeff:2009hq}. This model is equivalent to the three-dimensional Fierz-Pauli action for a massive spin-2	field at the linearized level.	In addition, NMG preserves parity symmetry which is not the case for the topological massive gravity (TMG) \cite{Deser:1982vy}.
	It has been shown that NMG admits
	BTZ and warped $AdS_3$ black holes solutions \cite{Donnay:2018ckb,Clement:2009gq}.
Due to different reasons three dimensional gravity has increasingly become eminent
since its black hole solution that has been known as
  BTZ black hole found by Bañados, Teitelboim, and Zanelli in 1992 \cite{DESER1984220,Edward Witten}.
The BTZ black hole in ``Schwarzschild'' coordinates is described by	the metric
\begin{eqnarray}
	ds^2 = -( N^\perp)^2dt^2 + f^{-2}dr^2
	+ r^2\left( d\phi + N^\phi dt\right)^2
	\label{a1}
\end{eqnarray}
where
\begin{eqnarray}
	N^\perp = f
	= \left( -M + {r^2\over\ell^2} + {J^2\over4r^2} \right)^{1/2} ,
	\quad N^\phi = - {J\over2r^2} \qquad  (|J|\le M\ell) .
	\label{a2}
\end{eqnarray}
BTZ black hole has two horizons and can  easily be found as
\begin{eqnarray}
r_\pm^2=\frac{M\ell^2}{2}\left(1\pm\left[1-\frac{J^2}{(M\ell)^2}\right]^{1/2}\right)~.
\end{eqnarray}
BTZ horizons are more obvious in the Eddington-Finkelstein-like coordinates,
\begin{eqnarray}
d v=dt+\frac{dr}{(N^\bot)^2}\,,\qquad\quad d\tilde{\phi}=d\phi-\frac{N^\phi}{(N^\bot)^2}dr~.
\end{eqnarray}
In this coordinate BTZ metric become simpler as:
\begin{eqnarray}
ds^2 = -( N^\perp)^2dv^2 +2dv dr
+ r^2\left( d\tilde{\phi} + N^\phi dv\right)^2\,.
\end{eqnarray}
In this coordinate horizons $r_{\pm}$ are located at $N^\bot=0$.

The above metric as the solution of Einstein-Hilbert action in three dimension, is expanded to the theory with higher curvature terms. This type of massive gravity for the first time is introduced by Bergshoeff, Hohm and Townsend (BHT). The action for the BHT massive gravity theory
is given by \cite{Deser:1982vy}
\begin{equation}
I_{BHT}=\frac{1}{16\pi G}\int d^{3}x\sqrt{-g}\left[  R-2\lambda-\frac{1}%
{m^{2}}K\right]  \ , \label{BHT action}%
\end{equation}
where $K$ stands for a precise combination of parity-invariant quadratic terms
in the curvature:%
\begin{equation}
K:=R_{\mu\nu}R^{\mu\nu}-\frac{3}{8}R^{2}\ . \label{K}%
\end{equation}
The field equations are then of fourth order and read%
\begin{equation}
G_{\mu\nu}+\lambda g_{\mu\nu}-\frac{1}{2m^{2}}K_{\mu\nu}=0\ , \label{feq}%
\end{equation}
where
\begin{equation}
K_{\mu\nu}:=2\nabla^{2}R_{\mu\nu}-\frac{1}{2}\left(  \nabla_{\mu}\nabla_{\nu
}R+g_{\mu\nu}\nabla^{2}R\right)  -8R_{\mu\rho}R_{\ \nu}^{\rho}+\frac{9}%
{2}RR_{\mu\nu}+g_{\mu\nu}\left[  3R^{\alpha\beta}R_{\alpha\beta}-\frac{13}%
{8}R^{2}\right]  \ .
\end{equation}	
One of the most interesting features of this theory has been mentioned in \cite{Bergshoeff:2009hq}. According to this paper, the \eqref{BHT action} action has constant curvature solution as
	\begin{equation}
R_{\alpha\beta}^{\mu\nu}=\Lambda_{\pm}\delta_{\alpha\beta}^{\mu\nu}~,
\end{equation}
where $\Lambda_{\pm}=2m(m\pm\sqrt{m^2-\lambda})$. Obviously this means that in the case of ${m^2=\lambda}$ theory has an unique maximally symmetric solution of fixed
curvature.

In the case of negative cosmological constant, $\Lambda=-\frac{1}{l^{2}}$ the solution of   \eqref{feq} field equation is \cite{Donnay:2020yxw,Banados:1992wn,Donnay:2018ckb}
	\begin{equation}
	ds^{2}=-\left(  \frac{r^{2}}{l^{2}}+br-\mu\right)  dt^{2}+\frac{dr^{2}}%
	{\frac{r^{2}}{l^{2}}+br-\mu}+r^{2}d\phi^{2}\ ,
	\label{Black hole negative lambda}%
	\end{equation}
	where b and $\mu$ are integration constants.
For the range of coordinates $-\infty<t<+\infty$, $0\leq\phi<2\pi$, it	describes asymptotically AdS black holes provided the lapse function $g_{tt}$	admits a positive real root.

	Finally, the solutions discussed in \eqref{Black hole negative lambda} can be generalized to the rotating case as \cite{Donnay:2020yxw,Oliva:2009ip}:	
	\begin{equation}
	ds_{a}^{2}=-N^2(r)f(r)dt^{2}+\frac{dr^{2}}{f(r)}+r^{2}\left(  d\phi+N^{\phi}dt\right)
	^{2}\ ,\label{Rotating BH}%
	\end{equation}
		with%
	\begin{align*}
	N &  =\left[  1-\frac{bl^{2}}{4\mathcal{G}}\left(  1-\Xi^{\frac{-1}{2}%
	}\right)  \right]  ^{2}\ ,\\
	N^{\phi} &  =-\frac{a}{2r^{2}}\left(  4GM-b\Xi^{\frac{-1}{2}}\mathcal{G}%
	\right)  \ ,\\
f &  =\frac{\mathcal{G}^{2}}{r^{2}}\left[  \frac{\mathcal{G}^{2}}{l^{2}}%
	+\frac{b}{2}\left(  1+\Xi^{\frac{-1}{2}}\right)  \mathcal{G}+\frac{b^{2}l^{2}%
	}{16}\left(  1-\Xi^{\frac{-1}{2}}\right)  ^{2}-4GM\ \Xi^{\frac{1}{2}}\right]
	\ ,
	\end{align*}
	and
	\begin{equation}
	\mathcal{G}=\left[  r^{2}-2GMl^{2}\left(  1-\Xi^{\frac{1}{2}}\right)
	-\frac{b^{2}l^{4}}{16}\left(  1-\Xi^{\frac{-1}{2}}\right)  ^{2}\right]
	^{\frac{1}{2}}\ .
	\end{equation}
	Here $\Xi:=1-a^{2}/l^{2}$, and the angular momentum is given by $J=Ma$, where
	$M$ is the mass (measured with respect to the zero mass black hole) and
	$-l<a<l$ is the rotation parameter. We can consider the following coordinate transformation:
	\begin{equation}
	dt   =dv-\frac{dr}{N\left(  r\right)  f\left(  r\right)  }\ ,\ \ \ \
	d\phi  =d\varphi+\frac{N^{\phi}\left(  r\right)  }{N\left(  r\right)f\left(r\right)
	}dr\ ,
	\end{equation}
	then, metric in \eqref{Rotating BH} can be written as follow
	\begin{equation}
	ds^{2}=-N^{2}\left(  r\right)  f\left(  r\right)  dv^{2}+2N\left(  r\right)
	drdv+ r^{2}  \left(  d\varphi +  N^{\phi}\left(
	r\right)    dv\right)  ^{2}\ .
	\label{BN4}
	\end{equation}
	In this work, we are going to consider symmetry properties of the BTZ black hole horizon through symmetry breaking formalism and Goldeston boson as well. This technique that firstly was suggested and developed by Nambu and Goldeston in the particle and condensed matter physics appears necessarily in models showing spontaneous breaking of continuous symmetries \cite{PhysRev.117.648}. Goldstone's theorem provides us with the chance to have Goldstone boson, as a massless scalar particle through global continuous symmetry breaking. Goldstone bosons that also can appear in the context of the general relativity play an important role in understanding behaviour of the system, especially in the low energy limit.
	
	\section{Near horizon symmetries of the massive gravity black hole}\label{Bibhas3}
	One of the most important concept in theoretical physics is symmetry and its conserve charge \cite{t,b,ds,s,adm,rt,wz,iw}. The symmetry in the gravitational theory is diffeomorphism. Finding the conserved charge and conserved current associated to these theories often is not as straightforward as others familiar field theories. However,  it must be considered that these concepts are vital to understand the thermodynamic properties of black holes. Recently, investigation of the black hole symmetries near the horizon has become popular and has studied widely \cite{Haco:2018ske,Haco:2019ggi,Setare:2019fxa}. Asymptotic symmetries are such that it preserves the form the metric near the asymptotic boundary or  near the horizon \cite{Maitra:2019eix,Maitra:2018saa}. To find out more about boundary symmetry we will follow the standard approach. This approach consider metric in the well known Gaussian null coordinate as
	\begin{eqnarray}
	ds^2=&&M(v,r,x^A) dv^2 + 2 dv dr + 2 h_A (v,r,x^A)dx^A dv
	\nonumber
	\\
	&&+ \mu_{AB} dx^A dx^B~,
	\label{eqn1}
	\end{eqnarray}
	In our case, the metric (\ref{BN4}) can be cast to the above form in the near horizon region when one uses the Gaussian null coordinates (see e.g. \cite{Donnay:2020yxw}). Here we shall perform the calculation in the original background (\ref{BN4}).  Note that three dimensions massive gravity case, $A=\phi$. The gauge conditions can be considered as
	\begin{eqnarray}
	\pounds_\zeta  g_{rr}= 0, \ \ \ \pounds_\zeta  g_{vr}=0, \ \ \ \pounds_\zeta  g_{Ar}=0~;
	\label{con}\\
	\pounds_\zeta  g_{vv} \approx \mathcal{O}(r); \ \ \pounds_\zeta  g_{vA} \approx \mathcal{O}(r); \ \ \  \pounds_\zeta  g_{AB} \approx \mathcal{O}(1)~.
	\label{con1}
	\end{eqnarray}
Some algebraic calculation leads to the diffeomorphism vectors associated to the above conditions \cite{Maitra:2019eix,Maitra:2018saa}:
\begin{eqnarray}
	&&\zeta^v = F(v,\phi)~;
	\\
	&&\zeta^r = - r \partial_v F- \partial_\phi F \int N^\phi(r) dr~;
	\nonumber\\
	&&\zeta^\phi = - \partial_\phi F \left(\frac{1}{r}-\frac{1}{r_+}\right)~.
	\nonumber
	\label{B}
	\end{eqnarray}
where $r_+$ is the black hole horizon. Under this diffeomorphism vector, the modified metric can be expressed as
\begin{eqnarray}
g'_{ab} &=& g^{(0)}_{ab} +h_{ab}
\label{rindler}
\end{eqnarray}
where $g^{(0)}_{ab}$ is given by (\ref{BN4}) and the components of the $h_{ab}$ perturbation are
	\begin{align}
	h_{vv}&=	- \bigl(N^\phi(r)\bigr)^2 r^2 - 2 \int N^\phi(r) dr N^\phi(r) r^2 \frac{\partial N^\phi(r)}{\partial r} \frac{\partial F(v, \phi)}{\partial \phi} \nonumber\\
	&+ 2 f(r) \int N^\phi(r) dr N(r) \frac{\partial N(r)}{\partial r} \frac{\partial F(v, \phi)}{\partial \phi} - 2 f(r) N(r) \bigl(N(r) + (r_{+}{} -  r) \frac{\partial N(r)}{\partial r}\bigr) \frac{\partial F(v, \phi)}{\partial v}\nonumber\\
	& + \bigl(N^\phi(r)\bigr)^2 r \bigl(r - 2 \int N^\phi(r) dr \frac{\partial F(v, \phi)}{\partial \phi} + 2 r_{+}{} \frac{\partial F(v, \phi)}{\partial v}\bigr) + \int N^\phi(r) dr N(r) \bigl(N(r) \frac{\partial f(r)}{\partial r} \frac{\partial F(v, \phi)}{\partial \phi}\nonumber  \\
	&- 2 \frac{\partial^{2}F(v, \phi)}{\partial v\partial \phi}\bigr)+ \frac{2 N^\phi(r) (r_{+}{} -  r) r \bigl(r_{+}{} r \frac{\partial }{\partial r} \frac{\partial F(v, \phi)}{\partial v} + \frac{\partial^{2}F(v, \phi)}{\partial v\partial \phi}\bigr)}{r_{+}{}} - 2 \frac{\partial^{2}F(v, \phi)}{\partial v^{2}}\nonumber \\
	&-  N(r) (r_{+}{} -  r) \bigl(N(r) \frac{\partial f(r)}{\partial r} \frac{\partial F(v, \phi)}{\partial v}\bigr)~;
	\end{align}
	\begin{align}
	h_{vr}&=-2 \bigl(N^\phi(r) + N(r) \frac{\partial \int N^\phi(r) dr}{\partial r} + \int N^\phi(r) dr \frac{\partial N(r)}{\partial r}\bigr) \frac{\partial F(v, \phi)}{\partial \phi}\\& + 2 (r_{+}{} -  r) \frac{\partial N(r)}{\partial r} \frac{\partial F(v, \phi)}{\partial v}\nonumber~;
	\end{align}
	\begin{align}
	\frac{r_+}{2}	h_{v\phi}&=- r_{+}{} f(r) \bigl(N(r)\bigr)^2 \frac{\partial F(v, \phi)}{\partial \phi}\nonumber \\&+ r_{+}{} \int N^\phi(r) dr \Bigl(- r \bigl(2 N^\phi(r) + r \frac{\partial N^\phi(r)}{\partial r}\bigr) \frac{\partial F(v, \phi)}{\partial \phi} -  N(r) \frac{\partial^{2}F(v, \phi)}{\partial \phi^{2}}\Bigr)\nonumber\\& + N^\phi(r) r \bigl(r_{+}{} N^\phi(r) r \frac{\partial F(v, \phi)}{\partial \phi} + (r_{+}{} -  r) \frac{\partial^{2}F(v, \phi)}{\partial \phi^{2}}\bigr) + r_{+}{} r \bigl(N^\phi(r) (2 r_{+}{} -  r)\nonumber\\& + (r_{+}{} -  r) r \frac{\partial N^\phi(r)}{\partial r}\bigr) \frac{\partial F(v, \phi)}{\partial v} + (r_{+}{} -  r) \bigl(r_{+}{} N(r) + r\bigr) \frac{\partial^{2}F(v, \phi)}{\partial v\partial \phi}~;
	\end{align}
	\begin{align}
	\frac{r_+}{2}h_{\phi\phi}=- r_{+}{} \int N^\phi(r) dr \frac{\partial F(v, \phi)}{\partial \phi} + r_{+}{} N^\phi(r) r \frac{\partial F(v, \phi)}{\partial \phi} + (r_{+}{} -  r) \bigl(\frac{\partial^{2}F(v, \phi)}{\partial \phi^{2}} + r_{+}{} \frac{\partial F(v, \phi)}{\partial v}\bigr)~;
	\end{align}
	\begin{align}
	h_{r\phi}=-2 \frac{\partial F(v, \phi)}{\partial \phi} + 2 N(r) \frac{\partial F(v, \phi)}{\partial \phi}~;
	\end{align}
	and finally
	\begin{align*}
	h_{rr}=0\,.
	\end{align*}

	In the next step, using the metric in \eqref{rindler} we will try to find the action for $F$. This is identified as the Goldstone field as it provides the change the macroscopic parameters of the black hole. The idea behind the construction of the action has been elaborated in \cite{Maitra:2019eix}. To summarize, note that the Lagrangian density changes by total derivative term of the form $\partial_a(\sqrt{-g}\mathcal{L}\zeta^a)$ under the diffeomorphism $x^a\rightarrow x^a+\zeta^a$ as it is diffeomorphism invariant. Here $\mathcal{L}$ is the Lagrangian. Now Gauss theorem reduces this total derivative change in the action as boundary term over the surface which encloses the manifold. Since in black hole spacetime a part of the boundary, which encloses the manifold volume, is horizon and as our analysis is valid on this boundary, we shall consider this part of the total derivative term as our aforesaid Lagrangian density. Having this idea the proposed required Lagrangian on horizon is of the form $\sqrt{-g}\mathcal{L}$. This will give the dynamics of Goldstone field $F$ (a detailed analysis has been provided in $II A$ of \cite{Maitra:2019eix}). Using this idea we write the required action as
	\begin{align}
	{\mathcal{S}} = \frac{1}{16 \pi G} \int d^2 x \sqrt{-g} \left(R-2\lambda-\frac{3}{8m^2}R^2+\frac{1}{m^2}R_{\mu\nu}R^{\mu\nu}\right)_{r=r_+}~;
	\end{align}
	Calculation of the above action under the corrected metric (\ref{rindler}) yields a result with several terms, accompanied by complicated constant coefficients. Our calculation leads to the following Lagrangian density:
	\begin{align}
	L&=A_1(\frac{\partial^{2}F(v, \phi)}{\partial v\partial \phi}\bigr)^2+A_2 \bigl(\frac{\partial^{2}F(v, \phi)}{\partial \phi^{2}}\bigr)^2+A_3\bigl(\frac{\partial F(v, \phi)}{\partial \phi}\bigr)^2
	\end{align}
	where the coefficients are given by
		\begin{align*}
		A_1&=
		\frac{6 }{M^2 r_+}
		\Bigl(-2 M^2 + 12 \bigl(N^\phi(r_{+}{})\bigr)^2 + 12 r_{+}{} N^\phi(r_{+}{}) \frac{\partial N^\phi(r_{+}{})}{\partial r_{+}{}} + 3 r_{+}{}^2 \bigl(\frac{\partial N^\phi(r_{+}{})}{\partial r_{+}{}}\bigr)^2\Bigr)~; 
		\end{align*}
		\begin{align*}
		A_2&=
		\frac{6 \bigl(N^\phi(r_{+}{})\bigr)^2}{M^2 r_{+}{}}
		\Bigl(-2 M^2 + 12 \bigl(N^\phi(r_{+}{})\bigr)^2 + 12 r_{+}{} N^\phi(r_{+}{}) \frac{\partial N^\phi(r_{+}{})}{\partial r_{+}{}} + 3 r_{+}{}^2 \bigl(\frac{\partial N^\phi(r_{+}{})}{\partial r_{+}{}}\bigr)^2\Bigr)~; 
		\end{align*}
		\begin{align}
		A_3&=\frac{1}{8M^2r_+}
		\Biggl(720 r_{+}{}^2 \bigl(N^\phi(r_{+}{})\bigr)^6 + 864 r_{+}{}^3 \bigl(N^\phi(r_{+}{})\bigr)^5 \frac{\partial N^\phi(r_{+}{})}{\partial r_{+}{}} + 144 r_{+}{} N^\phi(r_{+}{}) \bigl(\frac{\partial f(r_{+}{})}{\partial r_{+}{}}\bigr)^2 \frac{\partial N^\phi(r_{+}{})}{\partial r_{+}{}} \nonumber\\
		&+ 12 \bigl(\frac{\partial f(r_{+}{})}{\partial r_{+}{}}\bigr)^2 \Bigl(-2 M^2 + 3 r_{+}{}^2 \bigl(\frac{\partial N^\phi(r_{+}{})}{\partial r_{+}{}}\bigr)^2\Bigr) + 24 r_{+}{}^2 \bigl(N^\phi(r_{+}{})\bigr)^3 \frac{\partial N^\phi(r_{+}{})}{\partial r_{+}{}} \Bigl(-2 M^2 r_{+}{} + 24 \frac{\partial f(r_{+}{})}{\partial r_{+}{}}\nonumber\\
		& + 3 r_{+}{}^3 \bigl(\frac{\partial N^\phi(r_{+}{})}{\partial r_{+}{}}\bigr)^2\Bigr) + 72 r_{+}{} \bigl(N^\phi(r_{+}{})\bigr)^4 \Bigl(-2 M^2 r_{+}{} + 8 \frac{\partial f(r_{+}{})}{\partial r_{+}{}} + 5 r_{+}{}^3 \bigl(\frac{\partial N^\phi(r_{+}{})}{\partial r_{+}{}}\bigr)^2\Bigr)\nonumber \\
		&+ \bigl(N^\phi(r_{+}{})\bigr)^2 \biggl(4 M^2 r_{+}{}^2 \lambda \lambda + 144 \bigl(\frac{\partial f(r_{+}{})}{\partial r_{+}{}}\bigr)^2 - 12 M^2 r_{+}{}^4 \bigl(\frac{\partial N^\phi(r_{+}{})}{\partial r_{+}{}}\bigr)^2 + 9 r_{+}{}^6 \bigl(\frac{\partial N^\phi(r_{+}{})}{\partial r_{+}{}}\bigr)^4\nonumber \\
		&- 48 \frac{\partial f(r_{+}{})}{\partial r_{+}{}} \Bigl(2 M^2 r_{+}{} - 3 r_{+}{}^3 \bigl(\frac{\partial N^\phi(r_{+}{})}{\partial r_{+}{}}\bigr)^2\Bigr)\biggr)\Biggr)~.\label{a1-3} 
		\end{align}
	Following the Euler-Lagrangian equation in the second order as
	\begin{eqnarray}
	\frac{\partial L}{\partial F} - \partial_{\mu}(\frac{\partial L}{\partial (\partial_{\mu} F)}) + \partial_{\mu} \partial_{\nu} (\frac{\partial L}{\partial (\partial_{\mu} \partial_{\nu} F)}) =0 .\label{GenEH}
	\end{eqnarray}
	one can find out equation of motion. In this case, the result is found to be
	\begin{eqnarray}
		2A_1
		\partial^2_\phi\partial^2_v F(v, \phi)+A_2\partial^4_\phi F(v, \phi)-A_3\partial^2_\phi F(v, \phi)	=0~.
		\label{B2}
	\end{eqnarray}
	
	In the next section we will find the solution of this equation and will also provide a semi-classical quantum analysis.

\section{thermodynamics}\label{Bibhas4}
	Considering  the symmetry of the three dimensional massive gravity, we assume the following ansatz
	\begin{eqnarray}
	F(v,\phi)=\sum_n e^{-i n\phi}g_n(v)\label{ansatz}
	\end{eqnarray}
	where $g_n(v)$ is a function of $\nu$. Substitution of this in (\ref{B2}) we obtain that $g_n(v)$ obeys the below differential equation:
	\begin{align}
	-n^2\Big[2A_1\partial^2_v g_n(v)-(A_3+A_2 n^2)g_n(v)\Big]=-n^2\Big[\partial^2_v g_n(v)-V(r_+)^2 g_n(v)\Big]=0~,
	\label{eom}
	\end{align}
	where $V(r_+)^2=\frac{(A_3+A_2 n^2)}{2A_1}$ is the effective potential which provides a unstable solution for the \eqref{eom} equation as
	\begin{align}
	g(v)=\alpha_1 \exp{\left(V(r_+)v\right)}+\alpha_2  \exp{\left(-V(r_+)v\right)}+\alpha_3\delta_{n,0}~.
	\label{eom1}
	\end{align}
	In the above $\alpha_1, \alpha_2$ and $\alpha_3$ are arbitrary constants and would be determined by initial conditions. The last term is there when $n=0$. We shall see in later section that this $n=0$ mode provides the Hawking expression of horizon temperature for BTZ black hole. In fact, sensitivity and behaviour of these initial conditions are  controlled by Lyapunov exponent, $\lambda_L$, that is given by \cite{Shenker:2013pqa,Sekino:2008he,Morita:2019bfr}
	\begin{align}
\lambda_L={V(r_+)}\label{eq1}
	\end{align}	
	Having had deeper understanding of quantum gravity and gauge/gravity correspondence recently the idea of the maximal Lyapunov exponent was proposed \cite{Maldacena:1997re,Shenker:2013pqa,Morita:2018sen}. In these studies, authors considered a many body system and found out a upper limit for Lyapunov exponent as following:
		\begin{align}
\lambda_L\leq \frac{2\pi T}{\hbar}\label{eq2}
		\end{align}
	where T is system's temperature and $k_B=1$. In particular, for a field theory that has dual black hole geometry $\lambda_L= \frac{2\pi T}{\hbar}$ or $T=\frac{\hbar \lambda_L}{2\pi}$ \cite{Shenker:2013pqa,Sekino:2008he,Tsuji:2017fxs}. Using two equations in \eqref{eq1} and \eqref{eq2}, one can find easily a interesting relation
 between $T$ and inverse harmonic potential  and can help us to suggest that the rotated BTZ horizon carry entropy.
	
	In fact the equation for $g_n$ is similar to that for IHO. It is now well known that such system is unstable and provides quantum temperature at the quantum regime. There are several ways to look into this perspective, see for example \cite{Morita:2019bfr,Hegde:2018xub,Dalui:2019esx,Dalui:2020qpt,Subramanyan:2020fmx}. For completeness here we schematically mention one of the approaches which argues for thermalization of the IHO modes.
	The equation in \eqref{eom} can be expressed as a Schrodinger's like Hamiltonan \cite{Maitra:2019eix}
		\begin{align}
	H=-\frac{\hbar^2}{2}\frac{\partial^2}{\partial^2 g_n}-\frac{1}{2}V(r_+)^2g_n(v)^2\label{shc}
		\end{align}
		and then Schrodinger's wave equation is
				\begin{align}
			\frac{\hbar^2}{2}\frac{\partial^2\Phi}{\partial^2 g_n}-\frac{1}{2}V(r_+)^2g_n(v)^2\Phi=E\Phi~.
			\label{shc1}
				\end{align}
This equation vividly represents the degeneration of  all energy level $E_n$ associated with the different $g_n$ states that can help to establish a relation between harmonic potential and thermal behaviour of black hole \cite{Maitra:2019eix}. One of the most important features of the Schrodinger's wave equation is tunneling that according to the classical mechanics  will never  happen.
 The probability that particles  leak to the right side of the potential
 through the quantum tunneling can be calculated as \cite{Moore:1991zv}
 \begin{eqnarray}
 P_{T/R} = \frac{1}{e^{\frac{2 \pi}{\hbar} \sqrt{\frac{1}{V(r_+ )}} |E|} +1} = \frac{1}{ e^{\beta |E|} +1}~,
 \label{trans}
 \end{eqnarray}
 where $\beta$ is the identified inverse temperature.
So if a particle with negative energy $(E<0)$ collided with the potential barrier $V(r_+)$, it could penetrate the potential and appear on the right side. Since particle has  negative energy, this collision increases the amount of energy in left side of the potential which can be interpreted as a thermal radiation \cite{Morita:2019bfr}.
Finally, saturated bound of the \eqref{eq2} equation
leads to the minimum temperature of the rotated BTZ black hole  as
		\begin{eqnarray}
\beta=\frac{2\pi}{\hbar}\sqrt{\frac{1}{V(r_+)}}\longrightarrow T=\frac{\hbar}{2\pi}\sqrt{V(r_+)}=\frac{\hbar}{2\pi}\sqrt{\frac{(A_3+A_2 n^2)}{2A_1}}~.
		\end{eqnarray}
		This expression is purely quantum as it vanishes in the classical limit $\hbar\rightarrow 0$.
		So we see that the asymptotic symmetries can provide an explanation of the thermalization of black hole horizon.
		
\section{BTZ solution limit}
According to the Gauge/Gravity correspondence, a gravitational theory that is dual to  a field theory in the bulk must be compatible in three dimension. This fact emphasis importance of the three dimensional gravity and BTZ solution as well. BTZ solution can be obtained from \eqref{Rotating BH} metric by neglect from $b$, i.e, $b=0$. Not only does this limit simplify coefficient in \eqref{a1-3}, but also provide temperature of BTZ black hole in three dimensions. The three coefficients in \eqref{a1-3} in the $b=0$ limit are as follow:
		\begin{eqnarray}
A1&=&-\frac{12}{r_+}\\
A_2&=&-\frac{48G^2J^2}{r_+^5}\nonumber\\
A_3&=&-\frac{2\left(G^2J^2l^2+6r_+^4\right)}{l^4r_+^3}\,,\nonumber\label{btzlimit}
		\end{eqnarray}
		and the correspondence temperature can be read as:
\begin{eqnarray}
T=\frac{\hbar}{2\pi}\sqrt{\frac{(A_3+A_2 n^2)}{2A_1}}=\frac{\hbar}{2\pi}\sqrt{\frac{G^2J^2}{r_+^2}\left(\frac{1}{6l^2}+\frac{4n^2}{r_+^4}\right)+\frac{r_+^2}{l^4}}
\end{eqnarray}
and in particular, when we consider the ground state, i.e, $n=0$ the temperature become:
\begin{eqnarray}
T=\frac{\hbar}{2\pi}\sqrt{\frac{G^2J^2}{6r_+^2l^2}
+\frac{r_+^2}{l^4}}
\end{eqnarray}
which is the well known temperature of the BTZ black hole.

\section{conclusion and discussion}\label{Bibhas6}
Black hole physics is one of the most interesting aspects of the theoretical physics. Recently  several serious attempts have been made to reveal its nature, including black hole thermal behaviour. One approach that try to draw a clear picture of black hole  pursues black hole symmetries \cite{Wald:1984rg,wz,Haco:2018ske}. Another method is symmetry breaking \cite{Maitra:2019eix}. This method has been successful in the quantum field theory and Higgs boson as its outstanding prediction  has been found in the laboratory. Spontaneous symmetry breaking is the feature of global symmetries and so can occur in the theories, including gravity,  that show these symmetries. In this case symmetry property of the BTZ black hole drove from fall off and gauge conditions. Here we showed that, the approach adopted in \cite{Maitra:2019eix}, can be successfully implemented for BHT theory of gravitation and moreover again the Goldstone field, corresponding to BMS-like symmetry, can illuminate the thermalization of horizon.

In this regard, we now address the differences and similarities between the present analysis and the work which has been done in \cite{Donnay:2020yxw} where also BMS like symmetries near the horizon of the present black hole have been exploited to understand the horizon thermalization. Below we summarize them.
	\begin{itemize}
		\item In \cite{Donnay:2020yxw}, the supertranslation parameter $F$ has been taken as function of $\phi$ only. Whereas we considered much more general one as it is function of both timelike coordinate $v$ as well as $\phi$. Of course both the choices are consistent with the near horizon gauge and fall-off conditions (\ref{con}) and (\ref{con1}). This generalization, as already being described, helped us to raise the parameter as Goldstone field whose dynamical behaviour has been quite naturally emerged in our case.
		\item The earlier work \cite{Donnay:2020yxw} mainly studied the asymptotic charges near the horizon. On the contrary we analyzed the dynamics of identified Goldstone field to understand the thermodynamics of horizon. More specifically the behaviour of Goldstone modes have been scrutinized.
		\item The authors of \cite{Donnay:2020yxw} focused on the value of the charge mainly for the zero mode. They observed that this mode can reproduce the well known black hole entropy. This provides a hint whether the asymptotic symmetries have a role to understand the microscopic description of horizon thermalization. But what are the degrees of freedom (dof) and what should be their dynamics are still illusive. In this regard, it may be pointed out that these dof must be thermal in nature. Since the asymptotic symmetries has a connection to the horizon entropy, we were here interested to investigate the symmetry parameters, like $F$ more closely. We found that although these symmetries respect the near horizon structure of the metric, but modifies the macroscopic parameters of the black hole. This can be thought as a spontaneous breaking of symmetry and hence the parameter, like $F$ can be regarded as Goldstone field. With this our target were to study the dynamics of $F$ and to investigate whether the modes are thermal in nature. Interestingly, the modes capture a natural instability in the near horizon regime in the form of inverse harmonic oscillator. A semiclassical analysis revels that such instability naturally captures a thermalization at the quantum level. Actually each mode is thermal with a particular value for temperature. Contrary to \cite{Donnay:2020yxw} where the horizon entropy has been highlighted, here we are talking about the temperature. As each mode came out to be at a certain temperature and the analysis is based on a particular dynamics of $F$, we feel that the present analysis provides yet another robust claim in favour of asymptotic symmetries to revel the dof responsible for thermalization. In addition, contrary to \cite{Donnay:2020yxw}, our present analysis is ``dynamical'' in nature.
		\item It may be noted that in \cite{Donnay:2020yxw} the calculated charge is identified as $TS$, where $S$ is entropy of the black hole. In this identification the prerequisite is the explicit expressions for $T$ and $S$ must be known to us. Whereas in the present calculation we identified the temperature from ``first principle''. In this sense and since for both the calculations zero mode leads to the required results, we feel that these approaches can be complementary of each other.
	\end{itemize}
	Having stated the above differences and similarities we want to mention that although here we are dealing with similar set of asymptotic symmetries like in \cite{Donnay:2020yxw}, but their parameters are not exactly identical to ours. This is because our $F$ has dynamics as it is timelike coordinate dependent, in addition to dependencies on horizon coordinates. Moreover, we came up with the thermalization in a completely different way. Our approach was to directly see whether the aforesaid modes are thermal in nature which is completely a ``dynamical analysis''. This certain distinct approach must lead to more understanding about the role of asymptotic symmetries in describing black hole thermalization. Although the method is identical to \cite{Maitra:2019eix} which was confined to simplest scenarios (Rindler and Schwarzschild spacetimes) in four spacetime dimensions, this present analysis is done in different dimensions and gravitation theory where the black hole has intrinsic rotation. Therefore it shows the robustness of the current method and hence this may be a good candidate for further progress in other different black hole scenarios.
Another point that must be addressed is about Lyapunov exponent.
Maldacena, Shenker and  Stanford in their interesting paper \cite{Maldacena:1997re} discussed about butterfly effect in a many body system (system with  many freedom degrees). They argued there is a sharp limit on the rate of growth of chaos in thermal many body quantum
systems. This limit will be controlled by Lyapunov exponent. In fact, they have conjectured that any perturbation in a thermal many body system can not develop faster than  exponentially, with Lyapunov exponent, i.e $\lambda_L$.
We, as others, \cite{Maitra:2019eix,Maitra:2018saa} have used this result to calculate black hole rotated BTZ black hole through an inverse harmonic potential, instead of a many body potential. However, it has been  showed \cite{wiggins1990introduction} that a
many body system as a chaotic one behaviour is controlled by two important factor, firstly, hyperbolic fixed point
and secondly,  broken homo-clinic orbit. The equation of motion of a  particle  near the
hyperbolic fixed point might be effectively captured by
the one-dimensional particle motions in an inverse harmonic potential that a
many body system as a chaotic  \cite{wiggins1990introduction}. Finally we feel that the present analysis, over all, further talks in favour of symmetries to illuminate the thermal nature of black holes. Hopefully we will have deeper understanding of these Goldstone fields in future.

\end{document}